\documentclass[superscriptaddress,aps,prl,twocolumn,showpacs,preprintnumbers,amsmath,amssymb,floatfix,groupedaddress]{revtex4}

\usepackage{graphicx}
\usepackage{dcolumn}
\usepackage{bm}

\newcommand{\eop}{\mathcal{E}}

\begin{document}


\title{Universal Approach to Optimal Photon Storage in Atomic Media}

\author{Alexey V. Gorshkov}
\author{Axel Andr\'e}
\affiliation{Physics Department, Harvard University, Cambridge, 
Massachusetts 02138, USA}
\author{Michael Fleischhauer}
\affiliation{Fachbereich Physik, Technische Universit{\"a}t Kaiserslautern, 
67633 Kaiserslautern, Germany}
\author{Anders S. S{\o}rensen}
\affiliation{QUANTOP, Danish National Research Foundation Centre of Quantum Optics,
Niels Bohr Institute, DK-2100 Copenhagen {\O}, Denmark}
\author{Mikhail D. Lukin}
\affiliation{Physics Department, Harvard University, Cambridge, Massachusetts 02138, USA}

\date{\today}


\begin{abstract}
We present a universal physical picture for describing storage and retrieval of photon wave packets in a $\Lambda$-type atomic medium. This physical picture encompasses a variety of
different approaches to pulse storage ranging from adiabatic reduction of the photon
group velocity and pulse-propagation control via off-resonant Raman fields to photon-echo based techniques. Furthermore, we derive an optimal control strategy for storage and 
retrieval of a photon wave packet of any given shape. All these approaches, when optimized, yield identical maximum efficiencies, which only depend on the optical depth of the medium.
\end{abstract} 


\pacs{42.50.Gy, 03.67.-a, 32.80.Qk, 42.50.Fx}

\maketitle


In quantum networks, states are easily transmitted by photons, but the photonic states need to be stored locally to process the information. Motivated by this and other ideas from quantum information science, techniques to facilitate controlled
interactions between single photons and atoms are now being actively explored
\cite{kimblerempe,fleischhauer0002,lukin03,hau01,phillips01,polzik04,eisaman05,kuzmich05, hemmermanson, kozhekin00, nunn06, moiseev01, kraus06}. A 
promising approach to a matter-light quantum interface
uses classical laser fields to manipulate pulses of light
in optically dense media such as atomic gases \cite{fleischhauer0002,lukin03,hau01,phillips01,eisaman05,kuzmich05,polzik04, kozhekin00, nunn06, moiseev01} or impurities embedded in a solid state material
\cite{hemmermanson, kraus06}. The challenge is to map an incoming signal pulse into a long-lived
atomic coherence (referred to as a spin wave), so that it can be
later retrieved ``on demand" with the highest possible efficiency.
Using several different techniques, significant experimental progress towards this goal has been made recently \cite{polzik04,eisaman05,kuzmich05}. A central question that emerges from these advances is which approach represents the best possible strategy and how the maximum efficiency can be achieved. In this letter, we present a physical picture that unifies several different approaches to photon storage in $\Lambda$-type atomic media and yields the optimal control strategy. This picture is based on two key observations.
First, we show that the retrieval efficiency of any given stored spin wave depends only on the optical depth $d$ of the medium. Physically, this follows from the fact that the branching ratio between collectively enhanced emission into desired modes and spontaneous decay (with a rate $2 \gamma$) depends only on $d$. The second observation is that the optimal storage process is the time reverse of retrieval (see also \cite{moiseev01, kraus06}). This universal picture implies that the maximum efficiency is the same for all approaches considered and depends only on $d$. It can be attained by 
adjusting the control or the shape of the photon wave packet.

\begin{figure}[hb]
\includegraphics[scale = 0.19]{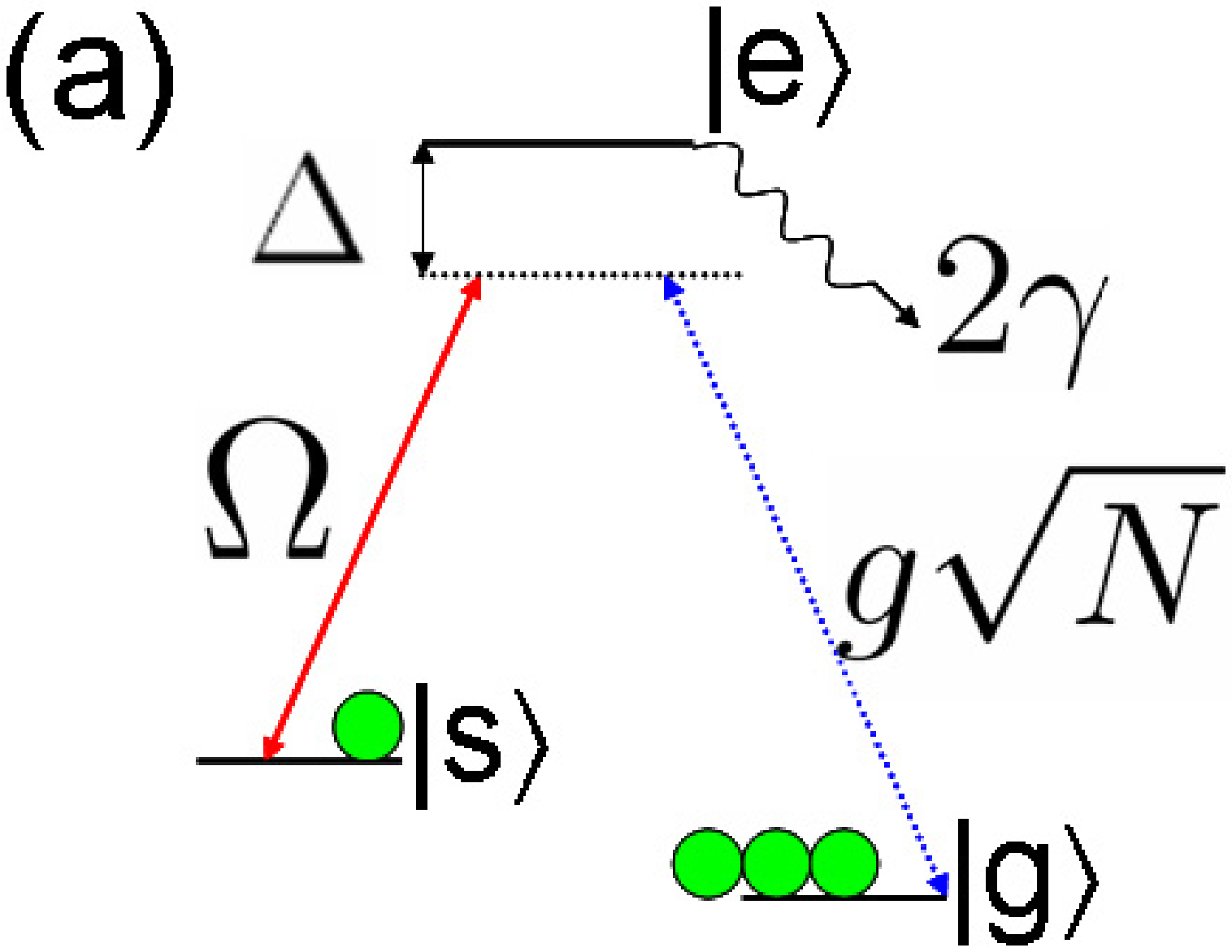}
\includegraphics[scale = 0.19]{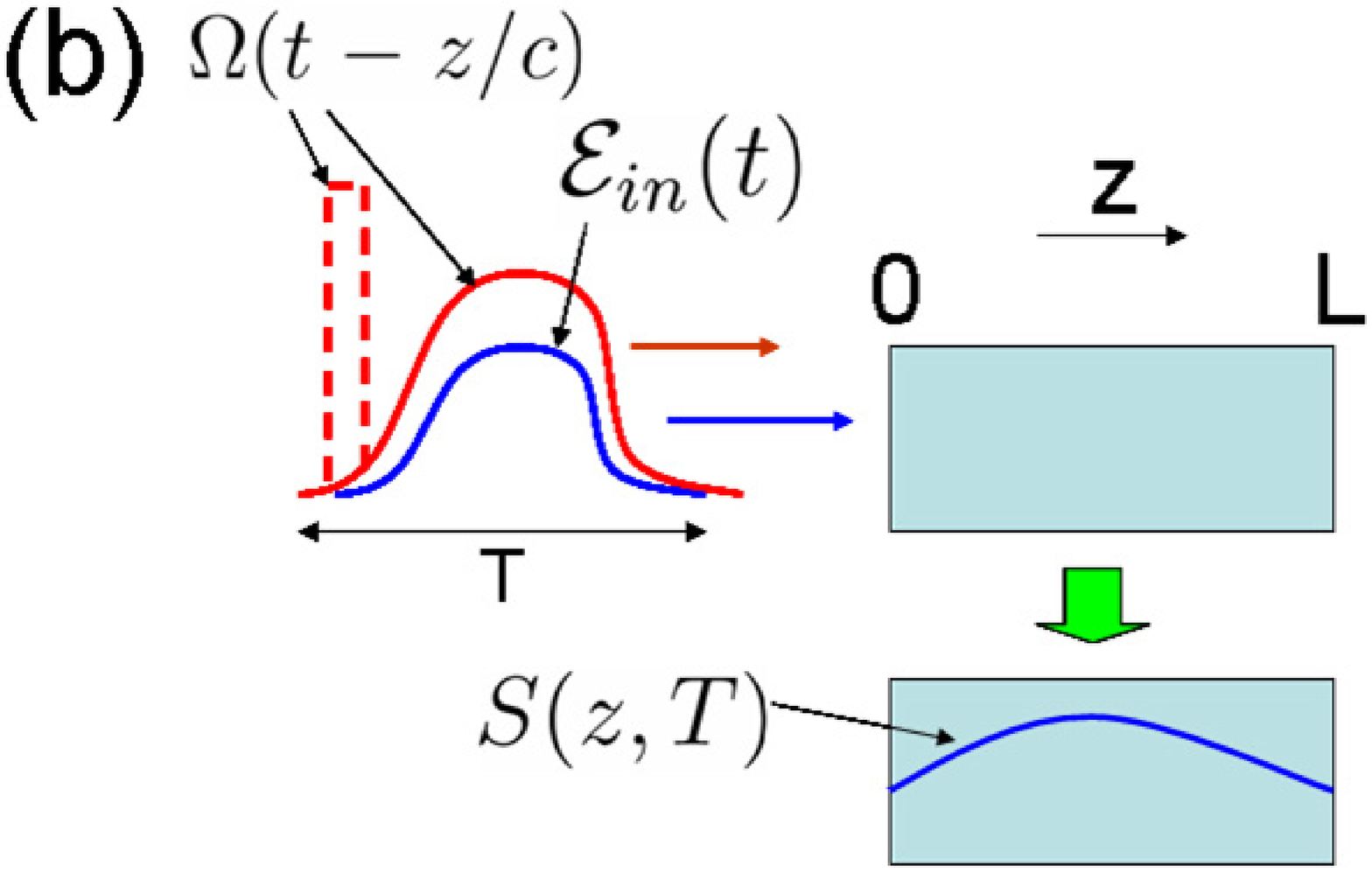}
\caption{(color online) (a) $\Lambda$-type medium coupled to a classical field with Rabi frequency $\Omega(t)$ and a quantum field with an effective coupling constant $g \sqrt{N}$. (b) Storage setup. The solid $\Omega$ curve is the generic control shape for adiabatic storage; the dashed line indicates a $\pi$-pulse control field for fast storage. For retrieval, the inverse operation is performed.\label{setup}}
\end{figure}

A generic model for a quantum memory uses the $\Lambda$-type level configuration shown in
Fig.\ \ref{setup}(a), in which a weak (quantum) signal field with freqeuncy $\nu$
is detuned by $\Delta$ from the $|g\rangle-|e\rangle$ transition. A  copropagating
(classical) control beam with the same detuning $\Delta$ from the $|s\rangle-|e\rangle$ transition is used to coherently manipulate the signal propagation and to facilitate the
light--atom state mapping. In this system several different  approaches to photon storage can be taken. In electromagnetically induced transparency
(EIT) \cite{fleischhauer0002,lukin03,hau01,phillips01,eisaman05,kuzmich05,hemmermanson},  resonant fields  ($\Delta=0$) are used to open a spectral transparency window, where the quantum field travels at a reduced group velocity, which is then adiabatically reduced to zero. In the Raman configuration \cite{kozhekin00,nunn06}, the fields have a large detuning   ($|\Delta| \gg \gamma d$) and  the photons 
are absorbed into the stable ground state $|s\rangle$ by stimulated Raman 
transitions. Finally, in the photon-echo approach \cite{moiseev01,kraus06}, photon storage is 
achieved by applying a fast resonant $\pi$-pulse, which maps excitations from the unstable excited state  $|e\rangle$ into the stable ground state $|s\rangle$. 

A common problem in all of these techniques is that the pulse should be  
completely localized inside the medium at the time of the storage.
For example, in the EIT configuration, a  
reduction in group velocity, which compresses the pulse to fit inside the medium, is accompanied by narrowing of the transparency window, which increases spontaneous emission.
Similarly, in the photon-echo technique, if a  
photon pulse is very short, its spectral width will be too large to be absorbed by the atoms.
To achieve the maximum storage efficiency
one thus has to make a compromise between the different sources of  
errors. Ideal performance is only achieved in the limit of infinite
$d$ \cite{fleischhauer0002}. 

In our model, illustrated in Fig.\ \ref{setup}(a), 
the incoming signal interacts with $N$ atoms in the uniform medium of length $L$ ($z=0$ to $z=L$) and cross-section area $A$. The control 
field is characterized by the slowly varying Rabi frequency $\Omega(t-z/c)$. 
$P(z,t)= \sqrt{N} \sum_i |g\rangle_i\langle e|/N_z$, where the sum is over all $N_z$ atoms in 
a small region positioned at $z$, describes the slowly varying collective 
$|g\rangle-|e\rangle$ coherence. All atoms are initially pumped into level $|g\rangle$. As indicated 
in Fig.\ \ref{setup}(b), we first map a quantum field mode with slowly varying envelope $\eop_\textrm{in}(t)$ (nonzero on $t \in [0,T]$ and incident in the forward direction at $z=0$) 
to some slowly varying mode of the collective $|s\rangle-|g\rangle$ coherence 
$S(z,t) = \sqrt{N} \sum_i |g\rangle_i\langle s|/N_z $. Then starting at a time 
$T_\textrm{r} > T$, we perform the inverse operation to retrieve $S$ 
back onto a field mode. As we explain below, the optimal efficiency is achieved by
sending the retrieval control pulse in the backward direction; storage followed by forward retrieval is, however, also considered. The goal is to solve the optimal control \cite{pontryagin86} problem of finding the control fields that will maximize the efficiency of storage followed by retrieval for given optical depth $d = g^2 N L/(\gamma c)$ and input mode $\eop_\textrm{in}(t)$. Here $c$ is the speed of light, the atom-photon coupling $g = \wp (\nu/(2 \hbar  \epsilon_0 A L))^{1/2}$ is assumed real for simplicity, and $\wp$ is the dipole matrix element. The efficiency is defined as the ratio of the number of retrieved photons to the number of incident photons. 

Since the quantum memory operates in the linear regime, an analysis of the
interaction process where all variables are treated as complex numbers 
is sufficient. In this limit the equations of motion read
\cite{paperII}
\begin{eqnarray} \label{eq1}
(\partial_t + c \partial_z) \eop(z,t) &=& i g \sqrt{N} P(z,t),
\\ \label{eq2}
\partial_t P(z,t) &=& - (\gamma + i \Delta) P(z,t) + i g \sqrt{N} \eop(z,t)+ 
\nonumber \\
& &+ i \Omega(t - z/c) S(z,t),
\\ \label{eq3}
\partial_t S(z,t) &=& i \Omega^*(t-z/c) P(z,t).
\end{eqnarray}
Here we have neglected the slow decay of $S$. For storage, the 
initial conditions are $\eop(0,t) = \eop_\textrm{in}(t)$, $\eop(z,0)=0$, $P(z,0)=0$, 
and $S(z,0)=0$. Being the shape of a mode, $\eop_\textrm{in}(t)$ is normalized according 
to $(c/L) \int_0^{T}  |\eop_\textrm{in}(t)|^2 d t = 1$, so the storage 
efficiency is given by $\eta_\textrm{s} = (1/L) \int_0^L |S(z,T)|^2 d z$. For the 
reverse process, i.e. retrieval, the initial conditions are $\eop(0,t) = 0$, $\eop(z,T_\textrm{r})=0$, 
$P(z,T_\textrm{r})=0$, and $S(z,T_\textrm{r})=S(L-z,T)$ for backward retrieval \cite{phasenote} 
or $S(z,T_\textrm{r})=S(z,T)$ for forward retrieval. The total efficiency in both cases is 
$\eta_\textrm{back/forw} = (c/L) \int_{T_\textrm{r}}^\infty |\eop_\textrm{out}(t)|^2 d t$, where $\eop_\textrm{out}(t) \equiv \eop(L,t)$. 

It is instructive to first discuss the retrieval process. In a co-moving frame $t' = t - z/c$, using a normalized coordinate $\zeta = z/L$ and a Laplace transformation in space $\zeta \rightarrow s$, Eq.\ (\ref{eq1}) gives $\eop(s,t') = i \sqrt{d \gamma L/c} P(s,t')/s$. Therefore, the retrieval efficiency is given by 
\begin{equation} \label{effeq}
\eta_\textrm{r} \!=\! \mathcal{L}^{-1} \bigg\{\gamma d/(s s') \int_{T_{\textrm{r}}}^\infty d t' P(s,t') \left[P(s'^*,t')\right]^*\!\bigg\},
\end{equation}
where $\mathcal{L}^{-1}$ means that two inverse Laplace transforms ($s \rightarrow \zeta$ and $s' \rightarrow \zeta'$) are taken and are both evaluated at $\zeta = \zeta' = 1$. To calculate $\eta_\textrm{r}$, we insert $\eop(s,t')$ found from Eq.\ (\ref{eq1}) into Eq.\ (\ref{eq2}) and use Eqs.\ (\ref{eq2},\ref{eq3}) to find  
\begin{eqnarray} \label{ddtPPstar}
\partial_t \left\{P(s,t') \left[P(s'^*,t')\right]^* + S(s,t') \left[S(s'^*,t')\right]^*\right\} 
\nonumber \\
= - \gamma (2+d/s+d/s') P(s,t') \left[P(s'^*,t')\right]^*. 
\end{eqnarray}
Eqs.\ (\ref{effeq},\ref{ddtPPstar}) allow us to express $\eta_\textrm{r}$ in terms of the initial and final values of the term inside the curly brackets in Eq.\ (\ref{ddtPPstar}). Assuming $P(s,\infty) = S(s,\infty)=0$ (i.e. no excitations are left in the atoms) and taking $\mathcal{L}^{-1}$, we get
\begin{eqnarray} \label{etar}
\eta_{\textrm{r}} = \int_0^1 d \zeta \int_0^1 d \zeta' k_{d}(\zeta,\zeta') S(\zeta,T_{\textrm{r}}) S^*(\zeta',T_{\textrm{r}}),  
\\
k_{d}(\zeta, \zeta') = \frac{d}{2} e^{-d \bigl(1-(\zeta+\zeta')/2\bigr)}
I_0\!\left(\!d \sqrt{(1-\zeta)(1-\zeta')}\right)\!,
\end{eqnarray}
where $I_0$ is the zeroth-order modified Bessel function of the first kind. Note that $\eta_\textrm{r}$ 
does not depend on $\Delta$ and $\Omega(t)$. Physically, this means that a fixed branching ratio exists between the transfer of atomic excitations into the output mode $\eop_\textrm{out}(t)$ and the decay into all other directions. This ratio only depends on $d$ and $S(\zeta,T_\textrm{r})$.

\begin{figure}[ht]
\includegraphics[scale = 0.205]{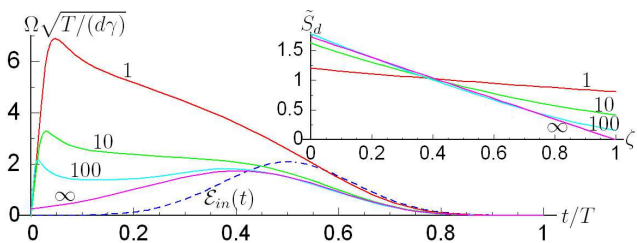}
\caption{(color online) Input mode $\eop_\textrm{in}(t)$ (dashed) and control fields $\Omega(t)$ (in units of $\sqrt{d \gamma/T}$) that maximize for this $\eop_\textrm{in}(t)$ the efficiency for resonant adiabatic storage (alone or followed by backward retrieval) at $d$ = 1, 10, 100, and $d \rightarrow \infty$. Inset: Optimal modes $\tilde S_d(1-\zeta)$ to retrieve from backwards at $d$ = 1, 10, 100, and $d \rightarrow \infty$ ($\zeta=z/L$). These are also normalized spin waves $S(\zeta,T)/\sqrt{\eta^\textrm{max}_\textrm{s}}$ in adiabatic and fast storage if it is optimized alone or followed by backward retrieval.\label{modecontrolswaves}}
\end{figure}

The efficiency $\eta_{\textrm{r}}$ in Eq.~(\ref{etar}) is an expectation value of a real symmetric operator $k_\textrm{r}(\zeta, \zeta')$ in the state $S(\zeta)$. It is, therefore, maximized when $S(\zeta)$ is the eigenvector (call it $\tilde S_d(\zeta)$) with the largest eigenvalue $\eta_\textrm{r}^\textrm{max}$ of the real eigenvalue problem
\begin{equation} \label{eigenproblem}
\eta_\textrm{r}\, S(\zeta) = \int_0^1 \!\!d \zeta'\,  k_{d}(\zeta,\zeta')\,  
S(\zeta').
\end{equation}
To find $\tilde S_d(\zeta)$, we start with a trial $S(\zeta)$ and iterate the integral in Eq.\ (\ref{eigenproblem}) several times. The resulting optimal spin wave
$\tilde S_d(1-\zeta)$ is plotted  in the inset of Fig.\ \ref{modecontrolswaves}
for $d = 1, 10, 100$, and $d \rightarrow \infty$.  These shapes represent a compromise attaining the smoothest possible spin wave with the least amount of (backward) propagation. 

We now discuss storage. We claim that if, for a given $d$, $\Delta$, and $\eop_\textrm{in}(t)$, we can find a control $\Omega(t)$ that retrieves backwards from $\tilde S_d(1-\zeta)$ into $\eop^*_\textrm{in}(T-t)$, then the time reverse of this control, $\Omega^*(T-t)$, will give the optimal storage of $\eop_\textrm{in}(t)$. To prove this, we represent our retrieval transformation as a unitary map $U[\Omega(t)]$ in the Hilbert space $\mathcal{H}$ spanned by subspace $A$ of spin-wave modes, subspace $B$ of output field modes, as well as a subspace containing (empty) input and reservoir field modes
(note that it is essential to include the reservoir modes, since the dynamics is unitary only in the full Hilbert space of the problem). For a given unit vector $|a\rangle$ in $A$ (a given spin wave), the retrieval efficiency is $\eta_\textrm{r}= |\langle b| U[\Omega(t)]|a\rangle|^2=|\langle a | U^{-1}[\Omega(t)]|b\rangle|^2$, where we have used the unitarity of $U[\Omega(t)]$, and where $|b \rangle$ is a normalized projection of $U[\Omega(t)]|a\rangle$ on $B$, i.e. the mode onto which the spin wave is retrieved. Introducing the time reversal operator $\mathcal{T}$ \cite{paperII}, we find $\eta_\textrm{r}=|\langle a |\mathcal{T} \mathcal{T} U^{-1}[\Omega(t)]\mathcal{T} \mathcal{T}|b\rangle|^2$. One can show \cite{paperII} that the time reverse of the inverse propagator $\mathcal{T} U^{-1}[\Omega(t)]\mathcal{T}$ is simply 
$U[\Omega^*(T-t)]$ so that we have $\eta_\textrm{r}=|\langle a |\mathcal{T} U[\Omega^*(T-t)]\mathcal{T} |b\rangle|^2$. This immediately tells us that the time-reversed control $\Omega^*(T-t)$ will map the time reverse of the retrieved pulse into the complex conjugate of the original spin-wave mode with the same efficiency. The optimal spin waves are, however, real so that complex conjugation plays no role. Furthermore, the storage efficiency cannot exceed $\eta_\textrm{r}^\textrm{max}$ since the time reverse of such storage would then by the same argument give a retrieval efficiency higher than $\eta_\textrm{r}^\textrm{max}$, which is a contradiction. Optimal storage is thus the time reverse of optimal backward retrieval and has the same efficiency $\eta^\textrm{max}_\textrm{s}=\eta^\textrm{max}_\textrm{r}$ (and involves the same optimal spin wave). 

To identify the input modes, for which the optimal storage can be achieved, we use Eqs.\ (\ref{eq1}-\ref{eq3}) to analytically solve the retrieval problem in two important limits: ``adiabatic" and ``fast". The ``adiabatic" limit, whose two special cases are the Raman and the EIT regimes discussed above, corresponds to a smooth control field, such that $P$ can be adiabatically eliminated in Eq.\ (\ref{eq2}). Using the Laplace transform technique to eliminate $\eop$ from Eqs.\ (\ref{eq1},\ref{eq2}), we reduce Eqs.\ (\ref{eq2},\ref{eq3}) to a simple differential equation on $S$. We solve it, compute $\eop$, and take the inverse Laplace transform to obtain  
\begin{eqnarray} \label{adiabaticoutput}
\eop_\textrm{out}(T_\textrm{r}\!\!+\!\!\frac{L}{c}\!+\!t) \!=\! - \sqrt{\frac{d \gamma L}{ c}} \!\int_0^1\!\!\!\! d \zeta 
\frac{\Omega(t)}{\gamma + i \Delta} 
e^{- \frac{\gamma d \zeta + h(t)}{\gamma + i \Delta}} 
 \nonumber\\
\times I_0 \left( 2 \sqrt{\gamma d \zeta h(t) }/\left(\gamma + i \Delta\right) \right) S(1-\zeta,T_\textrm{r}),
\end{eqnarray} 
where $h(t)=\int_0^t d t' |\Omega(t')|^2$. We will now show that for a given $d$, $\Delta$, and spin wave $S(\zeta)$, one can always find a control $\Omega(t)$ that maps $S(\zeta)$ to any 
desired normalized output mode $\eop_2(t)$ of duration $T_\textrm{out}$, so that  $\eop_\textrm{out}(T_\textrm{r}\!\!+\!\!\frac{L}{c}\!+\!t) = \sqrt{\eta_\textrm{r}} \eop_2(t)$ (provided we are in the ``adiabatic" limit $T_\textrm{out} d \gamma \gg 1$ \cite{paperII}). To do this, we replace $\eop_\textrm{out}(T_\textrm{r}\!\!+\!\!\frac{L}{c}\!+\!t)$ in Eq. (\ref{adiabaticoutput}) with $\sqrt{\eta_\textrm{r}} \eop_2(t)$, integrate the norm squared of both sides from $0$ to $t$, change variables $t \rightarrow h(t)$, and get
\begin{eqnarray}
&\eta_\textrm{r} \int_0^t d t' \left|\eop_2(t')\right|^2 \!=\! \frac{d \gamma L}{c} 
\int_0^{h(t)} d h' \big| \int_0^1 d \zeta 
\frac{1}{\gamma + i \Delta} 
e^{- \frac{\gamma d \zeta + h'}{\gamma + i \Delta}}& 
 \nonumber\\ \label{hequation}
&\times I_0 \left(2 \sqrt{\gamma d \zeta h' }/\left(\gamma + i \Delta\right) \right) S(1-\zeta,T_\textrm{r}) \big|^2,&
\end{eqnarray}
which allows us to solve numerically for the unique $h(t)$. Then $|\Omega(t)| = \left(\frac{d}{dt} h(t)\right)^{1/2}$, while the phase is found by inserting $h(t)$ into Eq.~(\ref{adiabaticoutput}).
Optimal storage controls then follow from the time reversal argument above. Fig.\ \ref{modecontrolswaves} shows a particular Gaussian-like input 
mode $\eop_\textrm{in}(t)$ and the corresponding optimal storage control shapes 
$\Omega$ \cite{controlsnote} for the case $\Delta = 0$ and $d = 1, 10, 100$, 
as well as the limiting shape of the optimal $\Omega$ as $d \rightarrow \infty$.
As we have argued, the normalized atomic mode $S(\zeta,T)/\sqrt{\eta^\textrm{max}_\textrm{s}}$, into which $\eop_\textrm{in}(t)$ is stored using these optimal control fields, is precisely $\tilde S_d(1 - \zeta)$, the optimal mode to retrieve backwards shown in the inset of Fig.\ \ref{modecontrolswaves}. 

The ``fast" limit corresponds to a short and powerful resonant retrieval control satisfying $\Omega \gg d \gamma$ that implements a perfect $\pi$-pulse between the optical and spin polarizations, $P$ and $S$. 
This retrieval and the corresponding storage technique are similar to the photon-echo method of Ref.\ \cite{moiseev01,kraus06}. Again using the Laplace transform technique, we find for a perfect $\pi$-pulse that enters the medium at time $T_\textrm{r}$
\begin{equation}\label{fastoutput}
\eop_\textrm{out}(T_\textrm{r}\!+\!\frac{L}{c}+t)\! =\! - \sqrt{\frac{\gamma d L}{c}}\!\! \int_0^1\!\!\!\! d \zeta 
e^{-\gamma t} J_0\!\! \left(\!2 \sqrt{{\gamma d \zeta t}} \right) 
\!S(1-\zeta,\!T_\textrm{r}),
\end{equation} 
where $J_0(x) = I_0(i x)$. Since the fast retrieval control cannot be shaped, at each $d$, there is, thus, only one mode (of duration $T \sim 1/(\gamma d)$) that can be stored optimally. This mode is the time reverse of the output mode in Eq. (\ref{fastoutput}) retrieved from the optimal spin wave $\tilde S_d$.  

We will now show that time reversal can not only be used to deduce optimal storage from optimal retrieval, but can also be used to find $\tilde S_d$ in the first place. In the discussion above,  the normalized projection of $U^{-1}|b\rangle$ on $A$ (call it $|a'\rangle$) might have a component orthogonal to $|a\rangle$. In this case, the efficiency of $U^{-1}$ as a map from $B$ to $A$ will be $\eta'_\textrm{r} = |\langle a' | U^{-1}|b\rangle| > \eta_\textrm{r}$. Now if the normalized projection of $U|a'\rangle$ on $B$ is not equal to $|b\rangle$, the map $U$ acting on $|a'\rangle$ will similarly have efficiency $\eta''_\textrm{r} > \eta'_\textrm{r} > \eta_\textrm{r}$. Therefore, such iterative application of $U$ and $U^{-1}$ converges to the optimal input in $A$ and the corresponding optimal output in $B$. 
Indeed, a detailed calculation \cite{paperII} shows that the search for the optimal spin wave by iterating Eq.\ (\ref{eigenproblem}) precisely corresponds to retrieving $S(\zeta)$ with a given control, time-reversing the output, and storing it with the time-reversed control profile.

\begin{figure}[t]
\includegraphics[scale = 0.09]{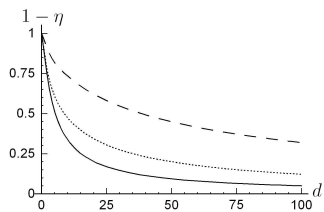}
\caption{$\eta^\textrm{max}_\textrm{back}$ (solid) and $\eta^\textrm{max}_\textrm{forw}$ (dotted) are maximum total efficiency for  
storage followed by backward or forward retrieval, respectively. $\eta_\textrm{square}$ (dashed) is the total efficiency for resonant storage of  
$\eop_\textrm{in}(t)$ from Fig.\ \ref{modecontrolswaves} followed by backward retrieval, where the storage control field is a na{\"i}ve square pulse. 
\label{effic}}
\end{figure}


This time-reversal optimization procedure for finding the optimal $|a\rangle \in A$ can be used to optimize not only retrieval, but also any map including storage followed by retrieval. For storage followed by backward retrieval, this procedure yeilds $\tilde S_d(1-\zeta)$ and maximum efficiency $\eta^\textrm{max}_\textrm{back}=(\eta^\textrm{max}_\textrm{r})^2$, since $\tilde S_d(1-\zeta)$ optimizes both storage and backward retrieval. Fig.\ \ref{effic} demonstrates that for resonant adiabatic storage of the field mode in Fig.\ \ref{modecontrolswaves} followed by backward retrieval, optimal controls result in a much higher 
efficiency $\eta^\textrm{max}_\textrm{back}$ than na{\"i}ve square control 
pulses on $[0,\! T]$ with power set by $v_g T = L$ ($\eta_\textrm{square}$ curve), 
where $v_g = c \Omega^2/(g^2 N)$ is the EIT group velocity \cite{lukin03}.

For the case of storage followed by forward retrieval, iterations yield the maximum efficiency $\eta^\textrm{max}_\textrm{forw}$ plotted in Fig.\ \ref{effic}. It is less than $\eta^\textrm{max}_\textrm{back}$ since with backward retrieval, storage and retrieval are each separately optimal, while for forward retrieval a compromise has to be made. From a different perspective, forward retrieval makes it more difficult to minimize propagation since the excitation has to propagate through the entire medium. 

In conclusion, we have shown that the performance of EIT, Raman, and photon-echo 
approaches to a quantum light-matter interface can be understood and optimized 
within a universal physical picture based on time reversal and a fixed branching 
ratio between loss and the desired quantum state transfer. For a given optical depth $d$, the 
optimal strategy yields a universal maximum efficiency and a universal optimal spin wave, thus, demonstrating a certain degree of equivalence between these three seemingly different approaches. We showed that the optimal storage can be achieved for any smooth 
input mode with $T d \gamma \gg 1$ and any $\Delta$ and for a class of resonant input modes satisfying $T d \gamma \sim 1$. The presented optimization of the storage and retrieval processes leads to a substantial increase in the memory efficiency. 
 
The results described here are of direct relevance 
to ongoing experimental efforts, where optical depth $d$ is limited by experimental constraints such as density of cold atoms, imperfect optical pumping, or competing nonlinear effects. For example, in two recent experiments \cite{eisaman05,kuzmich05}, $d \sim 5$ was used. $\eta^\textrm{max}_\textrm{back}$ and $\eta_\textrm{square}$ curves in Fig.\ \ref{effic} indicate that at this $d$, by properly shaping the control pulses, the efficiency can be increased by more than a factor of $2$. Direct comparison to experiment, however, will require the inclusion of decoherence processes and other imperfections. 
In Ref.\ \cite{paperII}, we discuss some of these imperfections, as well as the details of the present analysis and its extensions to atomic ensembles enclosed in a cavity and to inhomogeneously broadened media.


Finally, we note that the time-reversal based iterative optimization we suggest is not only a convenient mathematical tool but is also a readily accessible experimental technique for finding the optimal spin wave and optimal input-control pairs: one just has to 
measure the output mode and 
generate its time reverse. Indeed, our optimization procedure has been recently verified experimentally 
\cite{Novikova06}. We also 
expect this procedure to be applicable to the optimization of other linear quantum maps both within the field of light storage(e.g.~light storage using tunable photonic crystals \cite{Yanik04}) and outside of it.


We thank M.D.\ Eisaman for 
discussions. M.F. thanks the Harvard Physics Department and ITAMP for hospitality during his visit. This work was supported by the NSF, Danish Natural Science Research Council, DARPA, Harvard-MIT CUA, and Sloan and Packard Foundations.



\begin{thebibliography}{99}
\bibitem{phasenote} Can ignore spin wave momentum if $\sqrt{d} \gg w_{sg} L/c$ \cite{paperII}.
\bibitem{controlsnote} Although $\Omega(t\!=\!0)\!\rightarrow\! \infty$ for optimal $\Omega$, truncating it does not affect $\eta_\textrm{s}$ \cite{paperII}. Indeed, the requirement on the control pulse energy is $\gamma d \int\! d t \, \Omega^2 \gg |\gamma d + i \Delta|^2$ \cite{paperII}.
\end{thebibliography}
\end{document}